\documentclass[10pt,a4paper,abstract=on]{scrartcl}
\usepackage[utf8]{inputenc}
\usepackage[english]{babel}
\usepackage{amsmath}
\usepackage{amsfonts}
\usepackage{amssymb}
\usepackage{hyperref}
\usepackage{tikz}
\usepackage{color}
\usepackage[left=2cm,right=2cm,top=2cm,bottom=3cm]{geometry}
\title{The S-Matrix of superstring field theory}
\author{Sebastian Konopka}
\begin{document}

\hfill{LMU-ASC 40/15}
\vspace{3cm}

\begin{center}
{\LARGE \textbf{The S-matrix of Superstring Field Theory}}
\vspace{2cm}

{\Large Sebastian Konopka\footnote{sebastian.konopka@physik.uni-muenchen.de}}
\vspace{1cm}

{\large \textit{Arnold Sommerfeld Center, Ludwig-Maximilians Universit\"at, \\[.3em]
Theresienstra\ss{}e 37, D-80333 M\"unchen, Germany}}
\vspace{2cm}

\end{center}

We show that the classical S-matrix calculated from the recently proposed superstring field theories give the correct perturbative S-matrix. In the proof we exploit the fact that the vertices are obtained by a field redefinition in the large Hilbert space. The result extends to include the NS-NS subsector of type II superstring field theory and the recently found equations of motions for the Ramond fields. In addition, our proof implies that the S-matrix obtained from Berkovits' WZW-like string field theory then agrees with the perturbative S-matrix to all orders.
\newpage

\tableofcontents

\section{Introduction}
A field theoretical formulation of string theory can give valuable insight into a possible non-perturbative description of the moduli space of quantum string vacua. For the open bosonic string such a formulation was first described in light-cone gauge and later reformulated in covariant form \cite{Witten:1985cc}. The algebraic structures contained in the latter are still at the heart of any covariant string field theory in use today. Almost at the same time an analogous formalism for superstring field theory was proposed \cite{Witten:1986qs}. However, its construction was highly formal and turned out to give divergent results due to collisions of local operators on the world-sheet and therefore required regularization \cite{Wendt:1987zh}. The modified string field theory was proposed in \cite{Arefeva:1989cp,Preitschopf:1989fc} and dealt with the problem by using a modified kinetic term. But, it is not clear whether this field theory reproduces the correct particle spectrum.
Recently, a new regularization in terms of small Hilbert space fields and smeared picture changing operators was given \cite{SSFT1}. If the latter formulation defines a valid open superstring field theory, its S-matrix must necessarily coincide with the usual perturbative string S-matrix calculated in the formalism of picture changing operators \cite{Friedan:1985ey,Friedan:1985ge} or in terms of integrals over supermoduli space \cite{D'Hoker:1988ta,Witten:2012bh}. In this paper we prove such equivalence to the former formalism at tree-level or genus $0$.

Let us now outline the main ingredients of this proof for open string field theory.
At the perturbative level bosonic string field theory provides a definition of the Polyakov path-integral for arbitrary matter part with $c=26$. This means that its tree-level perturbation series gives rise to a regularized version of integrals over the whole moduli space of punctured discs. The Feynman perturbation series of planar tree-level diagrams in Siegel gauge coincides with the usual description of the color-ordered amplitude as an integral over the positions of all but three punctures. On the other hand the vertices of open string field theory satisfy the axioms of a cyclic $A_\infty$ algebra. At the algebraic level, the connection between the S-matrix and the $A_\infty$ algebraic structure is established through the so-called \emph{minimal model}. For any $A_\infty$ structure there exists an $A_\infty$ structure on the cohomology $H^\bullet(Q)$ in such a way that this induced structure is $A_\infty$-quasi-isomorphic to the original one. An explicit formula for the minimal model and the $A_\infty$-quasi-isomorphisms is formulated in terms of sums over all planar tree diagrams \cite{Kajiura:2003ax} and we argue that the matrix elements of the induced maps coincide with the color-ordered S-matrix.

The open superstring field theory action for the NS-sector in \cite{SSFT1} was found by requiring the vertices to be in the small Hilbert space and that they constitute a cyclic $A_\infty$ algebra. The solution was eventually obtained through a field redefinition in the large Hilbert space from a free theory. The field redefinition was constructed by integration of a pair of differential equations,
\begin{align*}
 \frac{\partial}{\partial t} \mathbf{M} &= [\mathbf{M}, \boldsymbol{\mu} ] \\
 \frac{\partial}{\partial s} \mathbf{M} &= [\boldsymbol{\eta}, \boldsymbol{\mu}],
\end{align*}
where $t$ was a deformation parameters, $s$ a formal parameter counting the so-called \emph{picture deficit} and $\boldsymbol{\mu}$ was an arbitrary function of $\mathbf{M}(s,t)$. 
In this paper, we show that at the level of the S-matrix this field redefinition leads to the needed insertions of picture changing operators (PCO) at the external legs. One essential feature of the proof is that it only requires the above two equations.
The proof itself is divided into three steps. In the first step we find an explicit expression of the minimal model using homological perturbation theory. Next, we argue that the products of the minimal model are identical to the perturbative, color-ordered $S$-matrix elements. We do this by showing that they satisfy a recursion relation that generates all planar tree diagrams. Finally, we evaluate the minimal model of open superstring field theory and relate it to the minimal model of the underlying bosonic string products. From which the postulated equivalence of the $S$-matrix of superstring field theory with the perturbative S-matrix calculated in the PCO formalism follows.

The outline of this paper is as follows: In section \ref{sec:sft} we review the reformulation of bosonic string field theory and open NS-superstring field theory from \cite{SSFT1} in homotopy algebraic language. In section \ref{sec:hpt}, we discuss some mathematical properties of the minimal model and find an explicit expression through an application of the homological perturbation lemma. Section \ref{sec:ssft} contains the key result of this paper. We apply the previously described techniques to evaluate the minimal model of open NS-superstring field theory. 
Quite interestingly, the proof can be adapted to all other superstring field theories based on homotopy algebraic methods. This includes the extension to the classical closed NS-NS superstring \cite{SSFT2}, the heterotic NS string \cite{SSFT2}, the equations of motion for the complete classical open superstring, closed superstring and heterotic string \cite{SSFT3}. As the arguments are very similar, we only discuss the extension to the closed NS-NS superstring in section \ref{sec:closed_ssft} and the extension to the equations of motion of the complete open superstring in section \ref{sec:ramond_ssft}. From there it should be clear that the extension to the remaining cases is straightforward. We also comment on the implications of our results on the S-matrix of Berkovits' WZW-like superstring field theory in section \ref{sec:berkovits}. Finally, we present our conclusions.

\section{Homotopy algebras and string field theory}
\label{sec:sft}
In this section we review the cobar description of the Chern-Simons-like formulation of the open NS-superstring given in \cite{SSFT1}. We start our short review with Witten's bosonic SFT and explain the modifications necessary to obtain a description of the open NS-superstring. Afterwards, we review the first quantized S-matrix and the concept of a minimal model associated to an $A_\infty$-algebra as well as its connection with the color-ordered $S$-matrix. 

\subsection{From Witten's bosonic OSFT to cyclic $A_\infty$ algebras}
The action of Witten's open bosonic string field theory \cite{Witten:1985cc} is formulated in terms of the world-sheet BPZ inner product $\langle \cdot, \cdot \rangle$, the world-sheet BRST operator $Q$ and a binary product $*$. These algebraic operations act on the Hilbert space of the underlying world-sheet CFT and form a differential graded algebra (DGA). Furthermore $\langle \cdot , \cdot \rangle$ is an invariant, graded-symmetric bilinear form of ghost number $-3$. The action of the bosonic string reads then as
\begin{align*}
	S &= \frac{1}{2} \langle \Phi, Q \Phi \rangle + \frac{g_o}{3} \langle \Phi, \Phi * \Phi \rangle,
\end{align*}
where $g_o$ denotes the open string coupling constant. In the following we set $g_o = 1$. The string field $\Phi$ is an element in the CFT Hilbert space at ghost number $1$. The action enjoys a gauge invariance,
\begin{align*}
	\delta \Phi &= Q \Lambda + \Lambda * \Phi - \Phi * \Lambda,
\end{align*}
where the gauge parameter $\Lambda$ is an arbitrary element in the CFT Hilbert space of ghost number $0$. It turns out \cite{Kajiura:2003ax} that gauge-invariance of $S$ is equivalent to $(Q,*)$ forming a DGA, i.e.~they verify the following axioms, $a,b,c \in \mathcal{H}$:
\begin{subequations}
\begin{align}
	Q^2 &= 0 \\
	Q ( a * b) &= Qa * b + (-1)^a\,a*Qb, \\
	(a*b)*c &= a*(b*c).
\end{align}
\label{eq:dga_axioms}
\end{subequations}
Here we introduced the symbol $(-1)^a$, which denotes the Grassmann parity of the state $a$ that is equal to the ghost number mod $2$. With these physical conventions $\Phi$ is an odd quantity. Let us further recall that $Q$ carries ghost number $1$ and $*$ carries no ghost number.

One way to study the space of DGAs (and $A_\infty$-algebras) on a fixed vector space is through their \emph{coalgebra representation}. First one performs a shift in the Grassmannality on the Hilbert space in such a way that all non-essential signs in (\ref{eq:dga_axioms}) are eliminated and then one treats the axioms as component equations on the tensor algebra $T\mathcal{H}$. Let us do this in greater detail, see \cite{SSFT1} for a more detailed review: We work with $\mathbb{Z}$-graded vector spaces so that we consider the same vector space but with different gradings as different objects. Given a $\mathbb{Z}$-graded vector space $\mathcal{H} = \bigoplus_{k\in\mathbb{Z}} \mathcal{H}_k$, we can define a new $\mathbb{Z}$-graded vector space by setting $(\mathcal{H}[1])_k = \mathcal{H}_{k+1}$. If we forget about the grading, $\mathcal{H}$ and $\mathcal{H}[1]$ are identical. The identity map $\mathbb{I}: \mathcal{H} \rightarrow \mathcal{H}$ can be lifted to a map $s: \mathcal{H}[1] \rightarrow \mathcal{H}$. $s$ is called the \emph{suspension} and carries degree $1$. The latter property is important as it tells us that $s$ is an anticommuting object. For example we have that $(s \otimes s)(s^{-1}\otimes s^{-1}) = - \mathbb{I} \otimes \mathbb{I}$.
Moreover, $s$ is invertible and we can introduce a new string field $\phi = s^{-1}\Phi \in \mathcal{H}[1]$. $\phi$ is even and carries degree $0$. As $s$ is invertible, we can express the DGA axioms equivalently on $\mathcal{H}[1]$,
\begin{align*}
	M_1 M_1 &= 0 \\
	M_1 M_2 + M_2(M_1\otimes\mathbb{I} + \mathbb{I}\otimes M_1) &= 0 \\
	M_2 (M_2\otimes\mathbb{I} + \mathbb{I}\otimes M_2) &= 0,
\end{align*}
where $M_1 = s^{-1} Q s$ and $M_2 = s^{-1} * (s\otimes s)$. We can think of these multilinear equations as linear equations on suitable tensor products of $\mathcal{H}[1]$. More precisely, we introduce linear maps on the tensor algebra $T(\mathcal{H}[1])$, $\mathbf{M}_k: T(\mathcal{H}[1]) \rightarrow T(\mathcal{H}[1]), k=1,2$ via
\begin{align}
	\mathbf{M}_k &= \sum_{r,s\geq 0} \mathbb{I}^{\otimes r} \otimes M_k \otimes \mathbb{I}^{\otimes s}.
	\label{eq:mk}
\end{align}
The DGA axioms (\ref{eq:dga_axioms}) can be written very compactly as
\begin{align*}
	\mathbf{M}^2 = 0 \mbox{ for }  \mathbf{M} &= \mathbf{M}_1 + \mathbf{M}_2.
\end{align*}
Linear combinations of maps of the form (\ref{eq:mk}) are known as coderivations and can be characterized in terms of a coalgebra structure on $T(\mathcal{H}[1])$. A coalgebra structure on a vector space $\mathcal{A}$ is specified by linear maps $\Delta: \mathcal{A} \rightarrow \mathcal{A} \otimes' \mathcal{A}$ and $\epsilon: \mathcal{A} \rightarrow \mathbb{C}$. These maps are called \emph{coproduct} and \emph{counit} respectively and are subject to the axioms
\begin{align*}
	(\Delta \otimes' \mathbb{I}_{\mathcal{A}}) \Delta &= (\mathbb{I}_{\mathcal{A}} \otimes' \Delta)\Delta \\
	(\epsilon \otimes' \mathbb{I}_{\mathcal{A}}) \Delta &= (\mathbb{I}_{\mathcal{A}}\otimes'\epsilon)\Delta = \mathbb{I}.
\end{align*}
On $T(\mathcal{H}[1])$ the coproduct and counit are defined as, $a_1, a_2, \ldots \in \mathcal{H}[1]$,
\begin{align*}
	\Delta (a_1 \otimes a_2 \otimes \cdots \otimes a_n) &= \sum_{r+s = n} (a_1 \otimes  a_2\otimes  \cdots\otimes  a_{r}) \otimes' (a_{r+1} \otimes a_{r+2}\otimes  \cdots\otimes  a_n), \\
	\epsilon (a_1 \otimes  a_2 \otimes \cdots\otimes  a_n) &= 0 \\
	\epsilon (1^*) &= 1,
\end{align*}
where the sum over $r$ and $s$ includes the values $r=0$ and $s=0$, in which case the corresponding product on the right has to be interpreted as $1^* \in \mathcal{H}^{\otimes 0} = \mathbb{C}\,1^* \cong \mathbb{C}$. Moreover, the tensor product on $T(\mathcal{H}[1])$ is written as $\otimes$, while the tensor product in $T(T(\mathcal{H}[1]))$ is denoted as $\otimes'$. A (first order) \emph{coderivation} $\mathbf{M}$ is then a linear map $\mathbf{M}: \mathcal{A} \rightarrow \mathcal{A}$ that fulfills a compatibility condition with $\Delta$
\begin{align*}
	\Delta \mathbf{M} &= (\mathbf{M} \otimes' \mathbb{I}_{\mathcal{A}} + \mathbb{I}_{\mathcal{A}} \otimes' \mathbf{M}) \Delta.
\end{align*}
It can be shown \cite{Kajiura:2003ax} that every coderivation can be written uniquely as a sum $\mathbf{M} = \sum_{k=0}^\infty \mathbf{M}_k$, where $\mathbf{M}_k$ is of the form (\ref{eq:mk}) with $M_k: \mathcal{H}[1]^{\otimes k} \rightarrow \mathcal{H}[1]$. See \cite{Erler:2015uba} for a detailed proof. We will call $M_k$ the $k$-string product in $\mathbf{M}$ or simply the $k$-product and write coderivations always in bold face. Another important property of coderivations is their closure under taking (graded) commutators. We will also need the notion of a \emph{cohomomorphism} $\mathcal{F}: \mathcal{A} \rightarrow \mathcal{A}'$ that is defined as a linear map intertwining the two coalgebra structures,
\begin{align*}
	\Delta' \mathcal{F} &= (\mathcal{F} \otimes' \mathcal{F}) \Delta \\
	\epsilon &= \epsilon' \mathcal{F}.
\end{align*}
Any cohomomorphism between tensor coalgebras $T(\mathcal{H}[1])$ and $T(\mathcal{H}'[1])$ is completely characterized by its projections $f_k = \pi_1 \mathcal{F} \iota_k: \mathcal{H}[1]^{\otimes k} \rightarrow \mathcal{H}'[1]$, where $\pi_k$ and $\iota_k$ are the canonical projections $\pi_k: T(\mathcal{H}[1]) \twoheadrightarrow (\mathcal{H}[1])^{\otimes k}$ and inclusion maps $\iota_k: (\mathcal{H}[1])^{\otimes k} \hookrightarrow T(\mathcal{H}[1])$. The most general form of a cohomomorphism is then given by \cite{Erler:2015uba}
\begin{align}
	\mathcal{F} &= \sum_{n=0}^\infty \sum_{r_1,r_2,\ldots,r_n} f_{r_1} \otimes f_{r_2} \otimes \cdots \otimes f_{r_n} = \sum_{n=0}^\infty (\pi_1 \mathcal{F})^{\otimes n}.
\label{eq:cohom}
\end{align}

With this new terminology we see that a DGA structure on $\mathcal{H}$ is equivalent to a degree $1$ coderivation $\mathbf{M}$ on $T(\mathcal{H}[1])$ that only has a non-trivial $1$- and $2$-product and is square zero $\mathbf{M}^2 = \frac{1}{2} [\mathbf{M},\mathbf{M}] = 0$. A square zero coderivation is also called a \emph{codifferential}.
From an algebraic point of view it seems very natural to relax the condition that $\mathbf{M}$ contains $1$- and $2$-products and just consider arbitrary degree $1$ codifferentials. We obtain a generalization of a DGA called a \emph{weakly homotopy associative algebra}. If the $0$-product (or tadpole) vanishes, i.e.~$M_0 = 0$, the algebraic structure is called a \emph{strongly homotopy associative algebra} or \emph{$A_\infty$ algebra}. The first few axioms of an $A_\infty$ algebra are
\begin{align*}
	M_1 M_1 &= 0 \\
	M_1 M_2 + M_2( M_1 \otimes \mathbb{I} + \mathbb{I} \otimes M_1) &= 0 \\
	M_1 M_3 + M_3( M_1 \otimes \mathbb{I}^{\otimes 2} + \mathbb{I} \otimes M_1 \otimes \mathbb{I} + \mathbb{I}^{\otimes 2} \otimes M_1) +2  M_2( M_2 \otimes \mathbb{I} + \mathbb{I} \otimes M_2) &= 0.
\end{align*}
The first two axioms are identical to the DGA axioms, but the associativity condition is only enforced up to $M_1$-exact terms. The conditions at higher order impose coherence conditions onto the higher products.

Witten's bosonic OSFT is also equipped with an invariant bilinear form $\langle \cdot , \cdot \rangle$. The invariance follows from the cyclicity of the action and states that for $a,b,c \in \mathcal{H}$,
\begin{align*}
	\langle Q a, b \rangle + (-1)^a \langle a, Qb \rangle &= 0 \\
	\langle a, b * c \rangle &=\langle a * b, c \rangle.
\end{align*}
In terms of elements in the suspended Hilbert space $\mathcal{H}[1]$ cyclicity then reads as, $\omega = \langle \cdot, \cdot \rangle s^{\otimes 2}$,
		\begin{subequations}
\begin{align}
	\omega( \mathbb{I} \otimes M_1 + M_1 \otimes \mathbb{I} ) &= 0 \\
	\omega( \mathbb{I} \otimes M_2 + M_2 \otimes \mathbb{I} ) &= 0.
\end{align}
	\label{eq:DGAcyclic}
	\end{subequations}
The bilinear map $\omega: \mathcal{H}[1] \otimes \mathcal{H}[1] \rightarrow \mathbb{C}$ is a graded antisymmetric map of degree $-1$ that is non-degenerate. Such a linear map is called a constant \emph{symplectic form}. The generalization of the cyclicity conditions (\ref{eq:DGAcyclic}) to arbitrary strongly homotopy associative algebras is now straightforward and the resulting algebraic structure is called a \emph{cyclic $A_\infty$ algebra}.
A rich source of cyclic $A_\infty$ algebras is provided by solutions to classical BV master equations, if functions on field space are required to be non-commutative but cyclic polynomials. Indeed, it can be shown that a cyclic $A_\infty$ structure is equivalent to a solution of a classical BV master equation $\{ S, S \} = 0$ \cite{Kajiura:2003ax}. It has been shown in \cite{Hata:1993gf} that for any cyclically invariant decomposition of the moduli space of punctured disks, the corresponding string field theory action satisfies a classical BV-master equation. Hence, the products $M_k$ together with the BPZ inner product define a cyclic $A_\infty$ algebra. For the bosonic open string infinitesimal deformations of the cyclic $A_\infty$ structure modulo field redefinitions are classified by physical closed string states \cite{Moeller:2010mh}, so that at least perturbatively the study of the cyclic BV-complex is equivalent to the world-sheet perspective.

In the rest of the paper we work in the suspended Hilbert space $\mathcal{H}[1]$ only and we drop the suffix $[1]$. Occasionally, we use $Q$ instead of $M_1$ to make some formulas look more familiar.

\subsection{Homotopy algebraic description of superstring field theory}
The main aim of \cite{SSFT1} has been to formulate a consistent field theory of interacting open superstrings in the small Hilbert space. However, due to presence of an additional conserved charge called \emph{picture} its formulation was not straightforward. Instead of constructing a field theory geometrically, a different approach was chosen: The primary goal was to find a non-trivial solution to the classical BV-master equation such that the string field lives in the small Hilbert space and carries ghost number $1$ and picture number $-1$. Moreover, it was required that the kinetic term be given by the world-sheet BRST operator $Q$ to ensure that the linearized equations of motion have the correct solution space. At the conceptual level, the formulation of classical perturbative open superstring theory in terms of picture changing operators (PCOs) was rewritten as a field theory. 
The complete answer was formulated as the result of a recursive algorithm, which takes as input the bosonic vertices together with a contracting homotopy of $[\boldsymbol{\eta},\cdot]$ acting on coderivations. In order to express the recursive structure most clearly, the so-called \emph{picture deficit}, a new grading on the space of coderivations was introduced: We say that a coderivation $\mathbf{M}_n$ has picture deficit $\mathrm{def}(\mathbf{M}_n) = n - 1 - \mathrm{pic}(\mathbf{M}_n)$, where $\mathrm{pic}(\mathbf{M}_n)$ denotes the picture number of $\mathbf{M}_n$. With this new grading, the restriction on picture number on the state space was lifted. A generic coderivation can be expanded in terms of picture deficit as
\begin{align*}
 \mathbf{M} &= \sum_{k=0}^\infty \mathbf{M}^{[k]},
\end{align*}
where $\mathrm{def}(\mathbf{M}^{[k]}) = k$. In order to keep track of the various homogeneous components $\mathbf{M}^{[k]}$, we introduced a formal variable $s$ of picture deficit $-1$ and used an identification $\mathrm{Coder}(T\mathcal{H}) \cong (\mathrm{Coder}(T\mathcal{H}) \otimes \mathbb{C}[s])^{[0]}$ under which
\begin{align*}
	\mathbf{M} &= \sum_{k = 0}^\infty s^k \mathbf{M}^{[k]}.
\end{align*}
The physical vertices that enter the action come then from a coderivation $\mathbf{M}^{[0]}$ with picture deficit $0$. By construction, the coderivation $\mathbf{M}$ has now picture deficit $0$. With these ingredients, the recursive algorithm is obtained from integrating the flow of a vector field $\delta = \frac{\partial}{\partial t}$ on the space of coderivations starting at the bosonic string products $\mathbf{M}_{0,k}$,
\begin{align}
\mathbf{M}|_{t=0} &= \mathbf{M}_0 = \mathbf{Q} + s \mathbf{M}_{0,2} + s^2 \mathbf{M}_{0,3} + \ldots \\
	\delta \mathbf{M} = \frac{\partial}{\partial t} \mathbf{M} &= [ \mathbf{M}, \boldsymbol{\mu} ]
	\label{eq:def1} \\
	\frac{\partial}{\partial s} \mathbf{M} &= [ \boldsymbol{\eta}, \boldsymbol{\mu}].
	\label{eq:def2}
\end{align}
$\boldsymbol{\mu}$ is an arbitrary coderivation solving the last equation. In \cite{SSFT1}, a particular contracting homotopy for $[\boldsymbol{\eta},\cdot]$ was used to solve the last equation for $\boldsymbol{\mu}$.
In \cite{SSFT1} it was shown that this recursive algorithm always yields the final string products in $\mathbf{M}^{[0]}$ up to a given number of inputs in a finite number of steps and that changing the duration $t$ of the flow yields is equivalent to a change in the coupling constant. We want to emphasize that our proof works for arbitrary families $\mathbf{M}(s,t)$ that solve equation (\ref{eq:def1}) and (\ref{eq:def2}) and does not depend on any particular choice of homotopy.

Let us review the particular contracting homotopy $\xi\circ$ for $[\boldsymbol\mu,\cdot]$ used in \cite{SSFT1} to solve equation (\ref{eq:def2}).
For unrelated reasons, it will also be useful (but not essential) in computing the $S$-matrix. Given that the cohomology of $\eta_0$ is trivial, we take a contracting homotopy $\xi$ (which must be of the sum of a BPZ-even operator and an $\eta$-closed BPZ-odd operator that we set to zero). For a coderivation $\mathbf{N}$ constructed from a single $k$-product $N_k$, we define $\xi \circ \mathbf{N}$ as 
\begin{align}
	\pi_1 (\xi \circ \mathbf{N}) \iota_k &= \frac{1}{k+1} \left(\xi N_k + (-1)^{|N|} \sum_{r+s=k-1} N_k (\mathbb{I}^{\otimes r} \otimes \xi \otimes \mathbb{I}^{\otimes s}) \right), \label{eq:specialxi}
\end{align}
and extend it linearly to arbitrary coderivations.
We also define another operation on coderivations, $X\circ$,
\begin{align*}
	X \circ \mathbf{N} &= [\mathbf{Q}, \xi \circ \mathbf{N} ] + \xi \circ [\mathbf{Q},\mathbf{N}].
\end{align*}
For the special homotopy $\xi\circ$ from equation (\ref{eq:specialxi}), it reads as
\begin{align*}
	\pi_1 (X \circ \mathbf{N}) \iota_k &= \frac{1}{k+1} \left(X N_k + \sum_{r+s=k-1} N_k (\mathbb{I}^{\otimes r} \otimes X \otimes \mathbb{I}^{\otimes s}) \right),
\end{align*}
where $X = [Q,\xi]$ is a picture changing operator used in \cite{SSFT1}.

\section{The minimal model}\label{sec:hpt}
\subsection{The minimal model of an $A_\infty$-algebra}\label{sec:minhpt}
The essential idea behind cohomology theories is that certain quantities interest can be represented in many different ways. Such quantities could be geometrical, topological invariants in mathematics or scattering matrix elements in physics. The various representations of that data are called models for the cohomology theory. When modelled with the help of dg-chain complexes, they typically include lots of auxilliary data and encode the physical information in the cohomology of some differential $Q$ together with some additional algebraic structure on that cohomology, like the gauge-invariant ``S-matrix''. The calculation of the gauge-invariant data living on the cohomology can be done in various models. Some of them lead to nice interpretations, while some of them allow for easy calculations. 

$A_\infty$ algebras are special cases of this idea: every $A_\infty$ algebra $\mathbf{M}$ induces an $A_\infty$-algebra structure $\mathbf{\tilde{M}}$ on the cohomology $H^\bullet(Q)$, the so-called \emph{minimal model}. We need a little more terminology. Given two $A_\infty$ algebras on $\mathcal{H}$ and $\mathcal{H}'$ described by coderivations $\mathbf{M}$ and $\mathbf{M}'$, we can define an \emph{$A_\infty$-morphism} $\mathcal{F}: (\mathcal{H},\mathbf{M}) \rightarrow (\mathcal{H}',\mathbf{M}')$ as a cohomomorphism $\mathcal{F}: T\mathcal{H} \rightarrow T\mathcal{H}'$ that intertwines both structures, $\mathcal{F} \mathbf{M} = \mathbf{M}' \mathcal{F}$. $\mathcal{F}$ is called an \emph{$A_\infty$-isomorphism} if it is invertible as a cohomomorphism. Let us denote by $f_k = \pi_1 \mathcal{F} \iota_k$ the component maps of $\mathcal{F}$. The component $f_1$ must the satisfy
\begin{align*}
  f_1 Q &= Q' f_1.
\end{align*}
Consequently $f_1$ is a chain map and, therefore, gives rise to a map $H^\bullet(Q) \rightarrow H^\bullet(Q')$. If the latter map is invertible, $\mathcal{F}$ is called an $A_\infty$-quasi-isomorphism. An $A_\infty$-algebra is called minimal if $Q = 0$.
The important \emph{minimal model theorem} states that any $A_\infty$-algebra is isomorphic to a minimal $A_\infty$-algebra and that this minimal model characterizes the $A_\infty$ algebra $\mathbf{M}$ completely, i.e.~any two $A_\infty$ algebras with $A_\infty$-isomorphic minimal models are quasi-isomorphic \cite{Kajiura:2003ax,Kontsevich:1997vb}. A nice review is \cite{2012arXiv1202.3245V}.

In the following we want to motivate the construction of the minimal model structure $\mathbf{\tilde{M}}$ via homological perturbation theory. In general there are two ways to define an algebraic structure on $H^\bullet(Q)$. The first approach takes arbitrary representatives for each cohomology class and defines a cohomology class by specifying some $Q$-closed vector. Then, one has to show that by changing the representative by a $Q$-exact piece modifies the answer only by a $Q$-exact piece. Alternatively, one can make a fixed choice of representative for each cohomology class and define the structure on them. The drawback of the latter construction is that it depends on the particular choice of representative and is not manifestly independent of it. Since one does not expect that the algebraic structures are independent of the choice of representative, one should at least require that two different choices give rise to isomorphic algebraic structures. In our case at hand, the first method only works for 2-product so that we need to resort to the second method for the higher products.

Let us now see how the first two products of the minimal model structure are constructed explicitly. In the first approach the induced $2$-product $\tilde{M}_2$ is obviously defined through the product $M_2$:
\begin{align*}
 \tilde{M}_2 &= M_2.
\end{align*}
This is a good choice because $Q$ is a derivation of this product so that the product of two $Q$-closed vectors is again $Q$-closed and shifting the representative by a $Q$-exact piece shifts the product by a $Q$-exact term. Thus, we have defined a binary product on cohomology $\tilde{M}_2: H^\bullet(Q) \otimes H^\bullet(Q) \rightarrow H^\bullet(Q)$. Since $M_2$ is associative up to homotopy, the induced product is completely associative. At this point we already have obtained an $A_\infty$-algebraic structure on the cohomology. Unfortunately, this new structure is not the minimal model because it is not $A_\infty$-quasi-isomorphic to the original structure in general.
Finding the $3$-product is a little bit more complicated. The naive guess $\tilde{M}_3 = M_3$ does not work, since it does not map $Q$-closed states into $Q$-closed states. Unless the induced $2$-product on cohomology is trivial, there is no way to define the $3$-product such that all $Q$-exact states decouple, so that the first method fails and we have to resort the the second. Choosing a representative for each cohomology class means that we select a section $i: H^\bullet(Q) \hookrightarrow \mathcal{H}$ of the canonical projection $p: \mathrm{ker}(Q) \subset \mathcal{H} \twoheadrightarrow H^\bullet(Q)$. Consequently we can find a homotopy $Q^\dagger$ such that we have
\begin{subequations}
  \begin{align}
    \mathbb{I} &= p \, i, \\
    \mathbb{I} - Q^\dagger Q - Q Q^\dagger &= i \, p \equiv P.
  \end{align}
  \label{eq:min_hom}
\end{subequations}
Our choice of binary product can then be expressed as
\begin{align*}
 \tilde{M}_2 &= p \,M_2\, i^{\otimes 2}.
\end{align*}
Using the $A_\infty$-relations together with (\ref{eq:min_hom}) one can show that this is indeed an associative product. Playing around a little bit, one discovers that the following product
\begin{align*}
 \tilde{M}_3 &= p\,\left( M_3 + M_2( (-Q^\dagger M_2) \otimes \mathbb{I} + \mathbb{I} \otimes (-Q^\dagger M_2) ) \right) i^{\otimes 3}
\end{align*}
satisfies $[\mathbf{\tilde{M}}_3,\mathbf{\tilde{M}}_2] = 0$. Hence, $\tilde{M}_2$ and $\tilde{M}_3$ satisfy the first two non-trivial $A_\infty$ relations.

As an example consider the DGA of differential forms on a compact manifold $X$, the so-called \emph{de Rham complex}. The cohomology theory that it models is purely topological and is known as the cohomology of $X$. The cohomology classes are in one-to-one correspondence with the homology $H_\bullet(X)$ through Poincar\'e duality. The induced associative product $\tilde{M}_2$ is then the cup product. If $X$ is orientable, we can also endow the DGA with an invariant symplectic form so that we obtain a cyclic DGA. The symplectic form is given by integration over $X$. In this case the product $\tilde{M}_2$ also calculates intersection numbers between cycles. The higher products $\tilde{M}_k, k\geq 3$ correspond to the Massey products and give refined topological information about $X$ \cite{2012arXiv1202.3245V}.

Guessing the higher order products is quite cumbersome and we want a systematic way to construct them. The answer is given in terms of the homological perturbation lemma \cite{HPT1}. Our goal is to construct $\mathbf{\tilde{M}}$ together with a pair of mutually inverse $A_\infty$-quasi-isomorphisms from/to the original $A_\infty$-structure. This means that we look for $A_\infty$-morphisms $\mathfrak{p}: (\mathcal{H},\mathbf{M}) \rightarrow (H^\bullet(Q),\mathbf{\tilde{M}})$ and $\mathfrak{i}: (H^\bullet(Q),\mathbf{\tilde{M}}) \rightarrow (\mathcal{H},\mathbf{M})$ such that $\mathfrak{p i} = \mathbb{I}_{T\mathcal{H}}$ and $\mathfrak{i p} \cong \mathbb{I}_{T\mathcal{H}}$, where the last requirement means that $\mathfrak{i p}$ is homotopic to $\mathbb{I}_{T\mathcal{H}}$, i.e.~there is a homotopy $H: T\mathcal{H} \rightarrow T \mathcal{H}$ such that
\begin{align*}
	\mathfrak{i p} &= \mathbb{I}_{T\mathcal{H}} + \mathbf{M} H + H \mathbf{M} \\
	\mathfrak{p} \mathbf{M} &= \mathbf{\tilde{M}} \mathfrak{p} \\
	\mathfrak{i} \mathbf{\tilde{M}} &= \mathbf{M} \mathfrak{i}.
\end{align*}
If $\mathbf{M} = \mathbf{Q}$ there are no induced products and the problem is easily solved in terms of $Q^\dagger$, $p$ and $i$ from (\ref{eq:min_hom}). The appropriate choices are
\begin{align*}
	H & = \sum_{r,s\geq 0} \mathbb{I}^{\otimes r} \otimes (-Q^\dagger) \otimes P^{\otimes s} \\
	\mathfrak{p} &= \sum_{r \geq 0} p^{\otimes r}, \\
	\mathfrak{i} &= \sum_{r \geq 0} i^{\otimes r} \\
	\tilde{\mathbf{M}} & = 0.
\end{align*}
They satisfy important compatibility conditions with the coalgebra structures $\Delta$ on $T\mathcal{H}$ and $T\mathcal{H}_p$,
\begin{subequations}
\begin{align}
	 \Delta\mathbf{Q} &= (\mathbf{Q} \otimes' \mathbb{I}_{T\mathcal{H}} + \mathbb{I}_{T\mathcal{H}} \otimes' \mathbf{Q}) \Delta \\
	 \Delta H &= (\mathbb{I}_{T\mathcal{H}} \otimes' H + H \otimes' \hat{P}) \Delta \\ \label{eq:homotopy}
	 \Delta\hat{P} &= (\hat{P} \otimes' \hat{P}) \Delta.
\end{align}
\label{eq:specialprop}
\end{subequations}
The homological perturbation lemma allows us to modify this solution of the ``free'' problem into a solution of the complete problem. 

Let $V$ be a graded vector space and $d$ some differential on $V$. The homological perturbation lemma \cite{HPT1} gives then a collection of formul\ae{} that allow for calculating the cohomology of a perturbed differential $d + \delta$, where $\delta$ is small in an appropriate sense. For our purposes one may think of $V = T\mathcal{H}$ and of $d$ as the codifferential $Q$ representing the free theory. We treat then the difference $\mathbf{M}_{\mathrm{int}} = \mathbf{M} - \mathbf{Q}$ as a perturbation in the sense of HPT. In order to state the perturbation lemma, we need a little bit more terminology. Let $(V,d)$ and $(W,D)$ be two chain complexes and let $\mathfrak{p}: V \rightarrow W$ and $\mathfrak{i}: W \rightarrow V$ be chain maps, s.t.~$\mathfrak{p i}= \mathbb{I}$ and $\hat{P} = \mathfrak{i p} = \mathbb{I} + h d + d h$ for some linear map $h: V \rightarrow V$ called the \emph{homotopy}. This collection of data $(V,W,d,D,\mathfrak{p},\mathfrak{i},h)$ is called \emph{homotopy equivalence data} and $(W,D)$ is said to be a \emph{deformation retract} of $(V,d)$. Let now $\delta$ be a perturbation of the chain complex $(V,d)$, i.e.~$(d + \delta)^2 = 0$. $\delta$ should be small in the sense that $(1 - \delta\,\, h)^{-1}$ exists. In our context $\delta = \mathbf{M}_{\mathrm{int}}$ represents the interaction part of the theory and is proportional to some coupling constant, so that the inverse exists at least perturbatively in the coupling constant. The main statement of the perturbation lemma is now that it is possible to deform the rest of the homotopy equivalence data in such a way that one retains valid homotopy equivalence data. More precisely,
\begin{subequations}
\begin{align}
	d' &= d + \delta \\
	\mathfrak{i}' &= \mathfrak{i} + h \delta (1- h \delta )^{-1} \mathfrak{i} \\
	\mathfrak{p}' &= \mathfrak{p} + \mathfrak{p} \delta (1- h\delta)^{-1} h \\
	D' &= D + \mathfrak{p}\delta (1-h\delta)^{-1} \mathfrak{i} \\
	h' &= h + h \delta (1-h \delta)^{-1}h\,.
\end{align}
\label{eq:hpt}
\end{subequations}
If the homotopy $h$ satisfies additional properties,
\begin{align}
	h \mathfrak{i} &= \mathfrak{p} h = h^2 = 0,
	\label{eq:sdr}
\end{align}
then $(W,D)$ is called a \emph{strong deformation retract} of $(V,d)$. Applying the homological perturbation lemma to the case at hand means replacing
\begin{align*}
	V &\rightarrow T\mathcal{H} & W &\rightarrow TH^\bullet(Q) \\
	d &\rightarrow \mathbf{Q} & D &\rightarrow 0 \\
	\delta &\rightarrow \mathbf{M}_{\mathrm{int}} & D' &\rightarrow \mathbf{\tilde{M}}.
\end{align*}
In particular, the new differential on $\mathbf{\tilde{M}}$ on $T H^\bullet(Q)$ reads as
\begin{align}
	\mathbf{\tilde{M}} &= \mathfrak{p} ( 1 - \mathbf{M}_{\mathrm{int}}H)^{-1} \mathbf{M}_{\mathrm{int}} \mathfrak{i}. \label{eq:minmodel}
\end{align}
Since $(TH^\bullet(Q),0)$ is a strong deformation retract of $(T\mathcal{H},\mathbf{Q})$, one can show that $\mathbf{\tilde{M}}$ is a coderivation and so defines a minimal $A_\infty$ structure on $H^\bullet(Q)$ and that $\mathfrak{p}$ and $\mathfrak{i}$ are cohomomorphisms. See appendix \ref{app:coh} for further details.

\subsection{The minimal model and Siegel gauge}
In string field theory the S-matrix is usually calculated in Siegel gauge. The various terms in the S-matrix calculated with the Siegel gauge propagator have a nice geometric interpretation in terms of disks obtained by glueing strips between the vertices and integrating over their lengths. A standard result in bosonic string field theory implies that one obtains the correct perturbative S-matrix \cite{Zwiebach:1990ba, Zwiebach:1992ie}. In this section we assume that the string background is flat Minkowski space or contains some uncompactified directions. The construction of the minimal model required the choice of contracting homotopy $Q^\dagger$ of the Hilbert space onto the cohomology. While such an operator always exists, it is not necessarily equal to the Siegel gauge propagator.
To see this, we make the choice $Q^\dagger = \frac{b_0}{L_0} (1 - e^{-\infty L_0})$. From this it follows that the physical projector $P$ is given by
\begin{align*}
	P = 1 - [Q,Q^\dagger] = e^{-\infty L_0}.
\end{align*}
Although $P$ is a projection operator, its image also contains unphysical states as $Q P \neq 0$ and $P$ is not a projection operator onto $H^\bullet(Q)$ but onto a larger vector space. We obtain a deformation retract of the original Hilbert space onto the image of $P$. At the end of section \ref{sec:minhpt} we argued that if we start with a strong deformation retract, we obtain an $A_\infty$ algebraic structure on $\mathcal{H}_p = P \mathcal{H}$ together with a pair of $A_\infty$-quasi-isomorphisms. Actually, we can relax these assumptions a little bit and require only that $Q^\dagger i = p Q^\dagger = (Q^\dagger)^2 = 0$. The differential on $\mathcal{H}_p$ is $Q P$. Application of the homological perturbation lemma means that we set
\begin{align*}
	V &\rightarrow T\mathcal{H} & W &\rightarrow T \mathcal{H}_p \\
	d &\rightarrow \mathbf{Q} & D &\rightarrow \mathbf{Q} \hat{P} \\
	\mathfrak{i} &\rightarrow \hat{P} & \mathfrak{p} &\rightarrow \hat{P} \\
	\delta &\rightarrow \mathbf{M}_{\mathrm{int}} & D' &\rightarrow \mathcal{S}(\mathbf{M}).
\end{align*}
This gives then the induced $A_\infty$-structure $\mathcal{S}(\mathbf{M})$ as
\begin{align}
	\mathcal{S}(\mathbf{M}) &= \mathbf{Q} \hat{P} + \hat{P}\mathbf{M}_{\mathrm{int}}   (1 - H \mathbf{M}_{\mathrm{int}})^{-1} \hat{P}.
	\label{eq:decmodel}
\end{align} 
We call $\mathcal{S}(\mathbf{M})$ the \emph{almost minimal model} and call the actual minimal model $\mathbf{\tilde{M}}$ the \emph{algebraic minimal model} whenever these distinctions are relevant. Since $\mathcal{S}(\mathbf{M})$ and $\mathbf{M}$ are $A_\infty$-quasi-isomorphic to each other, by the minimal model theorem the minimal model will be $\mathbf{\tilde{M}}$ in both cases. As discussed in the next section, the maps in $\mathcal{S}(\mathbf{M})$ are given by sums over planar tree-level diagrams with propagators $-\frac{b_0}{L_0}$. Calculation of the minimal model for $\mathcal{S}(\mathbf{M})$ requires us to choose a contracting homotopy for $QP$. The projector $P$ puts the states onto the mass-shell. This means that for operator $\mathcal{O}_1$ and $\mathcal{O}_2$ that are obtained by restricting space-time momentum preserving operators on $\mathcal{H}$ to operators on $\mathcal{H}_p$ we have the identity
\begin{align}
	\mathcal{O}_1 P \mathcal{O}_2 &= 0
	\label{eq:Pint}
\end{align}
for states with generic momentum. Thus, only diagrams with no internal lines contribute generically to the minimal model maps, but the only such diagrams are the vertices of $\mathcal{S}(\mathbf{M})$, which are identical to the perturbative $S$-matrix and coincide with the minimal model generically. Consequently, we can calculate the $S$-matrix either in Siegel gauge or using a complete gauge-fixing, but obtain the same answers.

\subsection{The minimal model and the $S$-matrix}
From a physical point of view the relevant information contained in the equations of motion are the observables and their expectation values. Observables are functions of the fields $\phi$, but we are not interested in arbitrary such functions, but only those that are gauge invariant. Moreover, we identify two gauge invariant functions if their difference vanishes on-shell, i.e.~is proportional to the equations of motion. The equivalence classes are the observables and can be thought of as functions on the moduli space of solutions modulo gauge-equivalence. The S-matrix measures then the obstruction for this moduli space to be smooth at $\phi = 0$ \cite{Cattaneo:2012qu}.
The most popular method for calculating the S-matrix perturbatively is through the use of Feynman diagrams. This approach, however, is not necessarily the best method for our purposes because the combinatorics for large $n$-point amplitudes is rather involved. Instead we use homological perturbation theory (HPT) which hides this difficulty and gives easy to manipulate formul\ae{} for the S-matrix. Using HPT to generate the full Feynman perturbation series is not new, see for example \cite{Albert:2008ui,Costello:caa,Gwilliam:2012jg}.

In the previous section we constructed the almost minimal model $\mathcal{S}(\mathbf{M})$ of $\mathbf{M}$. The claim is that its products are identical to the color-ordered $S$-matrix elements of a string field theory with vertices encoded in the codifferential $\mathbf{M}$. More precisely, we claim that the color-ordered S-matrix $S$ for $Q$-closed states $\Phi_i$ can be written as
\begin{align*}
	S(\Phi_1,\Phi_2,\ldots,\Phi_{n+1}) &= (-1)^{\Phi_{1}} \omega( \Phi_1 \otimes \pi_1 \mathcal{S}(\mathbf{M})_n(\Phi_2,\Phi_3,\ldots,\Phi_{n+1})  ),
	\end{align*}
	or more abstractly as
	\begin{align*}
	S &= \omega( \mathbb{I} \otimes \pi_1 \mathcal{S}(\mathbf{M}) ),
\end{align*}
where $\hat{P}$ is an extension of the physical projector to the tensor algebra, $\omega$ the symplectic form. The almost minimal model $\mathcal{S}(\mathbf{M})$ is gauge-invariant in the sense that
\begin{align}
	[Q, \mathcal{S}(\mathbf{M}) ] &= 0.
	\label{eq:Sgaugeinv}
\end{align}
Let us now justify these formulas.

The products in $\mathcal{S}(\mathbf{M})$ are identical to the color-ordered S-matrix elements of the underlying field theory $d+\delta$ since they are $A_\infty$-quasi isomorphic and thus have identical moduli spaces \cite{Kajiura:2003ax}. In the $L_\infty$-case this has been shown by Kontsevich in \cite{Kontsevich:1997vb} in his proof of deformation quantization. However, there is a more elementary way to verify this claim. The essential contribution to $\mathcal{S}(\mathbf{M})$ is given by
\begin{align}
	\hat{P} \mathbf{M}_{\mathrm{int}}  (1 - H \mathbf{M}_{\mathrm{int}})^{-1} \hat{P}.\label{eq:Smin}
\end{align}
We want to interpret this expression as a sum over planar rooted tree diagrams. To this end, we need to introduce a set of Feynman rules. A \emph{planar rooted tree diagram} is a planar graph of genus $0$ with a distinguished external line that we call its root. A Feynman diagram for states $\Phi_1,\Phi_2,\ldots,\Phi_n$ is obtained from a planar rooted tree with $n$ external lines as follows:
\begin{enumerate}
	\item Assign the state $(-1)^{\Phi_1} \omega( \Phi_1 \otimes \mathbb{I})$ to the root, $\Phi_2$ to the first leg next to the root in clockwise order and so on.
	\item To the vertex connected to the root assign the multilinear map $M_k$, where $k+1$ is the number of edges connected to it.
	\item To each other vertex of valence $k+1$ assign the operator $-Q^\dagger M_k$.
	\item Compose these multilinear maps according to the shape of the diagram.
\end{enumerate}
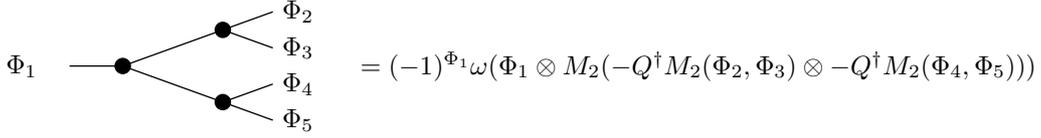
\begin{figure}[t]
	\begin{center}
		\begin{tikzpicture}
	\path
		(-1cm,0) node[anchor=east] {$\Phi_1$};		
		
	\draw[line width=.5pt] (-.7cm,0) -- (0,0);
	\draw[line width=.5pt] (0,0) -- (20:1.4cm);
	\draw[line width=.5pt] (0,0) -- (-20:1.4cm);
	\fill[color=black]
		(0,0) circle(1mm);
	\draw[line width=.5pt]
		(0,0) circle(1mm);
		
	\draw[line width=.5pt] (20:1.4cm) -- +(20:.7cm) node[anchor=west] {$\Phi_2$};
	\draw[line width=.5pt] (20:1.4cm) -- +(-20:.7cm) node[anchor=west] {$\Phi_3$};
	\draw[line width=.5pt] (-20:1.4cm) -- +(20:.7cm) node[anchor=west] {$\Phi_4$};
	\draw[line width=.5pt] (-20:1.4cm) -- +(-20:.7cm) node[anchor=west] {$\Phi_5$};
		
	\fill[color=black]
		(20:1.4cm) circle(1mm);
	\draw[line width=0.5pt]
		(20:1.4cm) circle(1mm);
	\fill[color=black]
		(-20:1.4cm) circle(1mm);
	\draw[line width=0.5pt]
		(-20:1.4cm) circle(1mm);
		
		\path (3cm,0)
			node[anchor=west] {$ = (-1)^{\Phi_1} \omega(\Phi_1 \otimes M_2( -Q^\dagger M_2(\Phi_2,\Phi_3) \otimes -Q^\dagger M_2(\Phi_4,\Phi_5) ) )$};
		\end{tikzpicture}
	
	\end{center}

	\caption{Illustration of the Feynman rules for five states. }
	\label{fig:diagrams}
\end{figure}

\begin{figure}[t]

\begin{center}
\begin{tikzpicture}
	\draw[line width=.5pt] (-1cm,0) -- (0,0);
	\draw[line width=.5pt] (0,0) -- (45:1cm);
	\draw[line width=.5pt] (0,0) -- (15:1cm);
	\draw[line width=.5pt] (0,0) -- (-15:1cm);
	\draw[line width=.5pt] (0,0) -- (-45:1cm);
	\fill[color=white]
		(0,0) circle(3mm);
	\draw[line width=0.5pt]
		(0,0) circle(3mm);

	\path
		(1.5cm,0) node {$=$};
		
	\begin{scope}[shift={(3cm,0)}]
	\draw[line width=.5pt] (-.7cm,0) -- (0,0);
	\draw[line width=.5pt] (0,0) -- (45:.7cm);
	\draw[line width=.5pt] (0,0) -- (15:.7cm);
	\draw[line width=.5pt] (0,0) -- (-15:.7cm);
	\draw[line width=.5pt] (0,0) -- (-45:.7cm);
	\fill[color=black]
		(0,0) circle(1mm);
	\draw[line width=.5pt]
		(0,0) circle(1mm);
			\path
		(1cm,0) node {$+$};
		\end{scope}
		
	\begin{scope}[shift={(5cm,0)}]
	\draw[line width=.5pt] (-.7cm,0) -- (0,0);
	\draw[line width=.5pt] (0,0) -- (30:1cm);
	\draw[line width=.5pt] (0,0) -- (0:1cm);
	\draw[line width=.5pt] (0,0) -- (-30:1cm);
	\fill[color=black]
		(0,0) circle(1mm);
	\draw[line width=.5pt]
		(0,0) circle(1mm);
		
	\draw[line width=.5pt] (30:1cm) -- +(20:.7cm);
	\draw[line width=.5pt] (30:1cm) -- +(-20:.7cm);
		
	\fill[color=white]
		(30:1cm) circle(3mm);
	\draw[line width=0.5pt]
		(30:1cm) circle(3mm);
			\path
		(2cm,0) node {$+$};
		\end{scope}
		
			\begin{scope}[shift={(8cm,0)}]
	\draw[line width=.5pt] (-.7cm,0) -- (0,0);
	\draw[line width=.5pt] (0,0) -- (30:1cm);
	\draw[line width=.5pt] (0,0) -- (0:1cm);
	\draw[line width=.5pt] (0,0) -- (-30:1cm);
	\fill[color=black]
		(0,0) circle(1mm);
	\draw[line width=.5pt]
		(0,0) circle(1mm);
		
	\draw[line width=.5pt] (0:1cm) -- +(20:.7cm);
	\draw[line width=.5pt] (0:1cm) -- +(-20:.7cm);
		
	\fill[color=white]
		(0:1cm) circle(3mm);
	\draw[line width=0.5pt]
		(0:1cm) circle(3mm);
			\path
		(2cm,0) node {$+$};
		\end{scope}
		
			\begin{scope}[shift={(11cm,0)}]
	\draw[line width=.5pt] (-.7cm,0) -- (0,0);
	\draw[line width=.5pt] (0,0) -- (30:1cm);
	\draw[line width=.5pt] (0,0) -- (0:1cm);
	\draw[line width=.5pt] (0,0) -- (-30:1cm);
	\fill[color=black]
		(0,0) circle(1mm);
	\draw[line width=.5pt]
		(0,0) circle(1mm);
		
	\draw[line width=.5pt] (-30:1cm) -- +(20:.7cm);
	\draw[line width=.5pt] (-30:1cm) -- +(-20:.7cm);
		
	\fill[color=white]
		(-30:1cm) circle(3mm);
	\draw[line width=0.5pt]
		(-30:1cm) circle(3mm);
			\path
		(2cm,0) node {$+$};
		\end{scope}
		
	\begin{scope}[shift={(3cm,-2cm)}]
	\draw[line width=.5pt] (-.7cm,0) -- (0,0);
	\draw[line width=.5pt] (0,0) -- (20:1cm);
	\draw[line width=.5pt] (0,0) -- (-20:1cm);
	\fill[color=black]
		(0,0) circle(1mm);
	\draw[line width=.5pt]
		(0,0) circle(1mm);
		
	\draw[line width=.5pt] (20:1cm) -- +(30:.7cm);
	\draw[line width=.5pt] (20:1cm) -- +(0:.7cm);
	\draw[line width=.5pt] (20:1cm) -- +(-30:.7cm);
		
	\fill[color=white]
		(20:1cm) circle(3mm);
	\draw[line width=0.5pt]
		(20:1cm) circle(3mm);
			\path
		(2cm,0) node {$+$};
		\end{scope}
		
			\begin{scope}[shift={(6cm,-2cm)}]
	\draw[line width=.5pt] (-.7cm,0) -- (0,0);
	\draw[line width=.5pt] (0,0) -- (20:1cm);
	\draw[line width=.5pt] (0,0) -- (-20:1cm);
	\fill[color=black]
		(0,0) circle(1mm);
	\draw[line width=.5pt]
		(0,0) circle(1mm);
		
	\draw[line width=.5pt] (-20:1cm) -- +(30:.7cm);
	\draw[line width=.5pt] (-20:1cm) -- +(0:.7cm);
	\draw[line width=.5pt] (-20:1cm) -- +(-30:.7cm);
		
	\fill[color=white]
		(-20:1cm) circle(3mm);
	\draw[line width=0.5pt]
		(-20:1cm) circle(3mm);
			\path
		(2cm,0) node {$+$};
		\end{scope}
		
			\begin{scope}[shift={(9cm,-2cm)}]
	\draw[line width=.5pt] (-.7cm,0) -- (0,0);
	\draw[line width=.5pt] (0,0) -- (20:1cm);
	\draw[line width=.5pt] (0,0) -- (-20:1cm);
	\fill[color=black]
		(0,0) circle(1mm);
	\draw[line width=.5pt]
		(0,0) circle(1mm);
		
	\draw[line width=.5pt] (20:1cm) -- +(20:.7cm);
	\draw[line width=.5pt] (20:1cm) -- +(-20:.7cm);
	\draw[line width=.5pt] (-20:1cm) -- +(20:.7cm);
	\draw[line width=.5pt] (-20:1cm) -- +(-20:.7cm);
		
	\fill[color=white]
		(20:1cm) circle(3mm);
	\draw[line width=0.5pt]
		(20:1cm) circle(3mm);
	\fill[color=white]
		(-20:1cm) circle(3mm);
	\draw[line width=0.5pt]
		(-20:1cm) circle(3mm);
		\end{scope}
	
\end{tikzpicture}
\end{center}

\caption{Graphical representation of equation (\ref{eq:proof.trees}) for four inputs. Big white circles represent components of $\pi_1 \Sigma$ and small solid circles elementary vertices $M_k$. Internal lines correspond to propagators $-Q^\dagger$. External lines pointing to the right represent physical states.}
\label{fig:trees}
\end{figure}
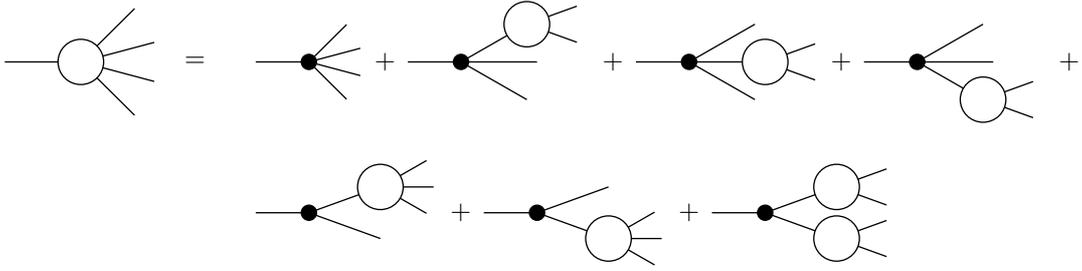

Figure \ref{fig:diagrams} illustrates this procedure using a particular example. With this new terminology we consider equation (\ref{eq:Smin}) and show that it is equal to the sum over all Feynman diagrams as just defined. 
Let us define two maps on the tensor algebra\footnote{At first one might think that one could replace $H \mathbf{M}_{\mathrm{int}}$ with $-Q^\dagger \mathbf{M}_\mathrm{int}$ in the next formula, but this leads to the wrong combinatorics. For example for the $5$-point function it generates too many tree diagrams with two $3$-vertices. The projections $P$ play an important role in establishing the identification with the $S$-matrix. },
\begin{align*}
	A &= (1 - H \mathbf{M}_{\mathrm{int}})^{-1} \hat{P} \\
	\Sigma &= \mathbf{M}_{\mathrm{int}} A.
\end{align*}
Notice that $\hat{P} \Sigma$ agrees with (\ref{eq:Smin}). The map $A$ is a cohomomorphism and, consequently, is determined by its component map $\pi_1 A: T\mathcal{H} \rightarrow \mathcal{H}$. Using the explicit form (\ref{eq:cohom}) of a cohomomorphism and the definition of the maps $M_k$ in equation (\ref{eq:mk}), one easily deduces the following pair of equations,
\begin{subequations}
\begin{align}
	\pi_1 A &= P + (-Q^\dagger) \pi_1 \Sigma \\
	\pi_1 \Sigma &= \sum_{k \geq 2} M_k (\pi_1 A)^{\otimes k} = \sum_{k \geq 2} M_k (P + (-Q^\dagger) \pi_1 \Sigma)^{\otimes k}.
\end{align}
\label{eq:proof.trees}
\end{subequations}
Equation (\ref{eq:proof.trees}) provides us with a recursive algorithm for the restrictions of $\pi_1 \Sigma$ to $n$ inputs. The reason being that the sum on the right hand side starts at $k = 2$ and so involves restrictions of $\pi_1 \Sigma$ to at most $n-1$ inputs. A graphical representation of equation (\ref{eq:proof.trees}) can be found in figure (\ref{fig:trees}). From the graphical representation it follows that $\pi_1 \Sigma$ is constructed from a sum over all planar tree-level Feynman diagrams. 
Equation (\ref{eq:proof.trees}) is recognized as the classical Dyson-Schwinger equation once one identifies $-Q^\dagger$ with the propagator and $M_k$ as interaction vertices of an action.

The important property (\ref{eq:Sgaugeinv}) that encodes the gauge-invariance of the S-matrix follows straightforwardly,
\begin{align*}
	[\mathbf{Q}, \mathcal{S}(\mathbf{M}) ] &= (1-\mathbf{M}_{\mathrm{int}} H)^{-1} \left( [\mathbf{Q},\mathbf{M}_{\mathrm{int}}H] \mathbf{M}_{\mathrm{int}} (1-H \mathbf{M}_{\mathrm{int}})^{-1} + [\mathbf{Q},\mathbf{M}_{\mathrm{int}}] \right) \notag \\
	&=  - \mathcal{S}(\mathbf{M}) \hat{P} \mathcal{S}(\mathbf{M}) = 0,
\end{align*}
where in the last step we used that fact that internal lines are generically off-shell.

Finally, we discuss the cyclicity properties of the S-matrix obtained from the almost minimal model for a cyclic $A_\infty$-structure. Since the operator $Q$ is BPZ-odd, any contracting homotopy $Q^\dagger$ can be written as the sum of a BPZ-even operator $\alpha$ and a BPZ-odd and $Q$-closed operator $\beta$, $[Q,\beta] = 0$. In order to calculate the deviation from cyclicity, we consider the sum,
\begin{align}
	\omega( \pi_1 \Sigma \otimes P + P \otimes \pi_1 \Sigma ) &= \sum_{k\geq 2} \omega\left( M_k (P + (-Q^\dagger)\pi_1\Sigma)^{\otimes k} \otimes P - M_k ( P \otimes (P + (-Q^\dagger) \pi_1 \Sigma)^{\otimes (k-1)}) \otimes (P + (-Q^\dagger) \pi_1 \Sigma ) \right) \notag\\
	&= - \sum_{k\geq 2} \omega\left( (-Q^\dagger) \pi_1 \Sigma \otimes \pi_1\Sigma + \pi_1 \Sigma \otimes (-Q^\dagger) \pi_1 \Sigma)\right) \notag\\
	&= 2 \omega( \beta (\pi_1 \Sigma) \otimes \pi_1 \Sigma).
	\label{eq:cyclic}
\end{align}
The S-matrix is manifestly cyclic provided $\beta = 0$. Moreover, if $\beta = [Q,R]$ for some operator $R$, the gauge-invariance (\ref{eq:Sgaugeinv}) tells us that the $S$-matrix is still cyclic, so that one can relax the requirement that $Q^\dagger$ has to be BPZ-even.

\section{Evaluation of the minimal model}
\label{sec:ssft}
\label{sec:proof}
In this section we apply the formalism for tree-level perturbation theory explained in the previous section to open superstring field theory. The original formulation can be found in \cite{SSFT1} and we review the relevant main ingredients and then express its S-matrix in terms of the usual perturbative S-matrix.
Equations (\ref{eq:def1}) and (\ref{eq:def2}) are the only two ingredients used in the evaluation of the almost minimal model along with a choice of propagator $Q^\dagger$. We claim that the following equation holds
\begin{align}
	X \circ \frac{\partial}{\partial s} \mathcal{S}(\mathbf{M}) - \frac{\partial}{\partial t} \mathcal{S}(\mathbf{M}) &= \left[ \boldsymbol{\eta}, \left[ \mathbf{Q}, \mathbf{T} \right] \right],
	\label{eq:main}
\end{align}
where $\mathbf{T}$ denotes some coderivation whose particular form is not relevant for the $S$-matrix. Before proving formula (\ref{eq:main}), we deduce the announced equivalence to the usual superstring S-matrix. To this end we recall that the $S$-matrix elements of our theory are calculated from the physical vertices $\mathbf{M}^{[0]}$, that is from the string products that are proportional to $s^0$. As the formula for the almost minimal model (\ref{eq:decmodel}) does not involve any operations on $s$, we conclude that the coefficient of $s^0$ in $\mathcal{S}(\mathbf{M})$, $\mathcal{S}(\mathbf{M})^{[0]}$, must be identical to the $S$-matrix of our field theory. Let us consider the $n$-product $\mathcal{S}(\mathbf{M})_n$. Since picture deficit is additive when composing multilinear maps, the highest power in $s$ of $\mathcal{S}(\mathbf{M})_n$ must be $s^{n-1}$. This is precisely the case when each vertex has maximal possible picture deficit. But these vertices are identical to the bosonic vertices, so that $\mathcal{S}(\mathbf{M})^{[n-1]}$ must calculate the bosonic $S$-matrix elements.
In order to convert the $\mathcal{S}(\mathbf{M})$ into a real S-matrix, we need to convert its output into an input using the symplectic form on the small Hilbert space $\omega_S = \omega( \mathbb{I} \otimes \xi )$, where the right-hand side is expressed in terms of the large Hilbert space symplectic form $\omega$. The S-matrix now reads as
\begin{align}
	S &= \omega( \mathbb{I} \otimes \xi \pi_1 \mathcal{S}(\mathbf{M}) ) = \sum_{k\geq 0} s^k S^{[k]}.
\end{align}
This linear functional has to be evaluated on vectors in the small Hilbert space that are generalized solutions to the equation $Q \phi = 0$. Moreover, this functional also has an expansion in terms of the variable $s$. Let us now look at the various coefficients of $s$ in (\ref{eq:main}). The right-hand side is a $Q$-exact and $\eta$-exact coderivation, so that we find on $H^\bullet(Q|\eta)$,
\begin{align*}
	\omega( \mathbb{I} \otimes \xi [\boldsymbol{\eta}, [\mathbf{Q},\mathbf{T}]]) &= \omega( \mathbb{I} \otimes [\mathbf{Q},\mathbf{T}]) = 0,
\end{align*}
where we used the compatibility of $\omega$ with both $Q$ and $\eta$. Thus, we can deduce from (\ref{eq:main}) the recursion relation, $k \geq 0$,
\begin{align}
	(k+1) S^{[k+1]} X &= \omega( \mathbb{I} \otimes \xi \,  \pi_1 \frac{\partial}{\partial t} \mathcal{S}(\mathbf{M})^{[k]} ) = \frac{\partial}{\partial t} S^{[k]}.
\end{align}
In order to solve this hierarchy of differential equations, we fix the number of external states to $n+1$ and consider $S_{n+1} = S \iota_{n+1}$ and we find that it satisfies the following differential equations,
\begin{align}
	\frac{\partial}{\partial t} S^{[0]}_{n+1} &= S^{[1]}_{n+1} X = \frac{1}{n+1} \sum_{r+s=n} S^{[1]}_{n+1} \left( \mathbf{1}^{\otimes r} \otimes X \otimes \mathbf{1}^{\otimes s} \right) \notag \\
	\frac{\partial}{\partial t} S^{[1]}_{n+1} &= 2 S^{[2]}_{n+1} X \notag \\
		&\vdots \notag \\
	\frac{\partial}{\partial t} S^{[n-1]}_{n+1} &= n S^{[n]}_{n+1} X\notag \\
	\frac{\partial}{\partial t} S^{[n]}_{n+1} &= 0.
	\label{eq:descent}
\end{align}
In the last equation we obtain zero, because all vertices have maximal possible picture deficit. These equations can be integrated, if we use the initial conditions $S^{[n]}_{n+1}(0) = S^{\mathrm{bos}}_{n+1}$ and $S^{[k]}_{n+1}(0) = 0$ for $k < n$, where $S^{\mathrm{bos}}_{n+1}$ denotes the bosonic $(n+1)$-S-matrix element. The result is,
\begin{align*}
	S^{[0]}_{n+1}(t) &= t^{n} S^{\mathrm{bos}}_{n+1} X^{n-1}.
\end{align*}
Moving around the picture changing operators $X$ does not change the S-matrix because of equation (\ref{eq:Sgaugeinv}). We can therefore distribute the PCOs such that each external leg has at most one $X$. Hence, the S-matrix $S^{[0]}$ can be calculated by taking all but two vertex operators in the $0$-picture and the remaining two in the $-1$-picture. The functional $S^{\mathrm{bos}}$ then inserts these vertex operators at the boundary of a disc and integrates over the possible positions, essentially by the validity of the usual bosonic string field theory construction. Consequently, $S^{[0]}$ is identical to the perturbative string S-matrix, as claimed.

It remains to prove (\ref{eq:main}). Before we consider the completely general case, we concentrate on the $3$-product in (\ref{eq:main}) for which the proof can be carried out by hand. In this case $\mathcal{S}(\mathbf{M})_3$ reads as
\begin{align*}
	\mathcal{S}(\mathbf{M})_3 &= \hat{P} ( \mathbf{M}_3 + \mathbf{M}_2( -Q^\dagger \mathbf{M}_2 \otimes \mathbf{1} + \mathbf{1} \otimes -Q^\dagger \mathbf{M}_2)) \hat{P}^{\otimes 3}.
\end{align*}
Note that is equation is between polynomials in $s$, so that it actually represents multiple equations. We can apply two operators, $\frac{\partial}{\partial t}$ and $\frac{\partial}{\partial s}$ to this equality,
\begin{align*}
	\frac{\partial}{\partial t} \mathcal{S}(\mathbf{M})_3 &= \hat{P} \left( \frac{\partial}{\partial t} \mathbf{M}_3 + \frac{\partial}{\partial t} \mathbf{M}_2( -Q^\dagger \mathbf{M}_2 \otimes \mathbb{I} + \mathbb{I} \otimes -Q^\dagger \mathbf{M}_2) + \mathbf{M}_2( -Q^\dagger \frac{\partial}{\partial t} \mathbf{M}_2 \otimes \mathbb{I} + \mathbb{I} \otimes -Q^\dagger \frac{\partial}{\partial t} \mathbf{M}_2) \right) \hat{P} \\
	\frac{\partial}{\partial s} \mathcal{S}(\mathbf{M})_3 &= \hat{P} \left( [\boldsymbol\eta, \boldsymbol\mu_3] + [\boldsymbol\eta, \boldsymbol\mu_2] ( -Q^\dagger \mathbf{M}_2 \otimes \mathbb{I} + \mathbb{I} \otimes -Q^\dagger \mathbf{M}_2) +   \mathbf{M}_2 ( -Q^\dagger [\boldsymbol\eta, \boldsymbol\mu_2]\otimes \mathbb{I} + \mathbb{I} \otimes -Q^\dagger [\boldsymbol\eta, \boldsymbol\mu_2]) \right) \hat{P} \\
	&= \hat{P} \left(  [\boldsymbol\eta, \boldsymbol\mu_3 + \boldsymbol\mu_2( -Q^\dagger \mathbf{M}_2 \otimes \mathbb{I} + \mathbb{I} \otimes -Q^\dagger \mathbf{M}_2) +  \mathbf{M}_2( -Q^\dagger \boldsymbol\mu_2 \otimes \mathbb{I} + \mathbb{I} \otimes -Q^\dagger \boldsymbol\mu_2)] \right) \hat{P}.
\end{align*}
In order to proceed, we apply $[\mathbf{Q}, \xi \circ ]$ to the second equality and use the property $[\boldsymbol{\eta}, \xi\circ] + \xi \circ [\boldsymbol\eta,\cdot] = \mathbb{I}$ to find,
\begin{align*}
	[\mathbf{Q}, \xi \circ \frac{\partial}{\partial s} \mathcal{S}(\mathbf{M}) ]_3 &= [\mathbf{Q}, [\boldsymbol\eta, \cdots]] + \hat{P} [\mathbf{Q},\boldsymbol\mu_3 + \boldsymbol\mu_2( -Q^\dagger \mathbf{M}_2 \otimes \mathbb{I} + \mathbb{I} \otimes -Q^\dagger \mathbf{M}_2) +  \mathbf{M}_2( -Q^\dagger \boldsymbol\mu_2 \otimes \mathbb{I} + \mathbb{I} \otimes -Q^\dagger \boldsymbol\mu_2)] \hat{P} \\
	&= [\mathbf{Q},[\boldsymbol\eta,\cdots]] + \frac{\partial}{\partial t} \mathcal{S}(\mathbf{M})_3,
\end{align*}
where the ellipsis corresponds to some irrelevant terms contributing to $\mathbf{T}$. Rearranging the latter equation a little bit yields,
\begin{align*}
X \circ \frac{\partial}{\partial s} \mathcal{S}(\mathbf{M})_3 - \frac{\partial}{\partial t} \mathcal{S}(\mathbf{M})_3 &= [\mathbf{Q},[\boldsymbol\eta,\cdots]].
\end{align*}
During the calculation we made use of equation (\ref{eq:Pint}), which allows us to drop terms involving the physical projector $P$ between operators when evaluating this expression on physical states, because generically the internal lines will be off-shell. Thus, for $n=3$, (\ref{eq:main}) follows.

The general case can be derived analogously. The starting point is equation (\ref{eq:decmodel}). The homotopy $H$ is constructed from the propagator $-Q^\dagger$ and from the physical projector $P$ via formula (\ref{eq:homotopy}). It follows then straightforwardly that
\begin{align}
	\frac{\partial}{\partial t} \mathcal{S}(\mathbf{M}) &= \hat{P} ( 1 - \mathbf{M}_{\mathrm{int}} H)^{-1} \frac{\partial}{\partial t} \mathbf{M} H   ( 1 - \mathbf{M}_{\mathrm{int}} H)^{-1} \mathbf{M}_{\mathrm{int}} \hat{P} + \hat{P} ( 1 - \mathbf{M}_{\mathrm{int}} H)^{-1}  \frac{\partial}{\partial t} \mathbf{M} \hat{P} = \notag \\ &= \hat{P} ( 1 - \mathbf{M}_{\mathrm{int}} H)^{-1}  [\mathbf{M},\boldsymbol\mu]  H   ( 1 - \mathbf{M}_{\mathrm{int}} H)^{-1} \mathbf{M}_{\mathrm{int}} \hat{P} + \hat{P} ( 1 - \mathbf{M}_{\mathrm{int}} H)^{-1}  [\mathbf{M},\boldsymbol{\mu}] \hat{P} = \notag \\
	&=  \hat{P} ( 1 - \mathbf{M}_{\mathrm{int}} H)^{-1}  [\mathbf{M},\boldsymbol{\mu}]  ( 1 - H \mathbf{M}_{\mathrm{int}})^{-1} \hat{P} \\
	\frac{\partial}{\partial s} \mathcal{S}(\mathbf{M}) &= \hat{P} ( 1 - \mathbf{M}_{\mathrm{int}} H)^{-1} \frac{\partial}{\partial s} \mathbf{M} ( 1 - H \mathbf{M}_{\mathrm{int}})^{-1}\hat{P}  = \hat{P}  ( 1 - \mathbf{M}_{\mathrm{int}} H)^{-1} [ \boldsymbol\eta, \boldsymbol\mu ]  ( 1 - H \mathbf{M}_{\mathrm{int}})^{-1} \hat{P}  \notag \\
	&= [\boldsymbol\eta, \hat{P} ( 1 - \mathbf{M}_{\mathrm{int}} H)^{-1} \boldsymbol\mu ( 1 - H \mathbf{M}_{\mathrm{int}})^{-1} \hat{P}  ] \equiv [\boldsymbol\eta, \boldsymbol{\rho} ], \label{eq:svardec}
\end{align}
where in the last step we used the fact that the interaction term, the physical projector and the homotopy $H$ commute with the coderivation $\boldsymbol\eta$. Note that $\boldsymbol\rho$ is a coderivation. Now, we solve (\ref{eq:svardec}) for $\boldsymbol\rho$ modulo $\eta$-exact terms using the contracting homotopy $\xi\circ$. We find that
\begin{align}
	{\boldsymbol\rho} &= \xi\circ \frac{\partial}{\partial s} \mathcal{S}(\mathbf{M}) + [\boldsymbol\eta, \xi \circ {\boldsymbol\rho}].
	\label{eq:rho}
\end{align}
In order to produce a PCO instead of a $\xi$ on the right-hand side, we calculate the commutator with the coderivation $\mathbf{Q}$. Calculating the commutator of $\mathbf{Q}$ with operators of the form $(1 - A)^{-1}$ is easy, once one recognizes that the Leibniz rule for $[\mathbf{Q},\cdot]$ implies that $[\mathbf{Q}, (1-A)^{-1}] = (1-A)^{-1} [\mathbf{Q},A] (1-A)^{-1}$. In our case we have $A = \mathbf{M}_{\mathrm{int}} H$ and, hence,
\begin{align*}
	[\mathbf{Q}, {\boldsymbol\rho}] &= \hat{P} (1-\mathbf{M}_{\mathrm{int}} H)^{-1} [\mathbf{Q},\,\mathbf{M}_{\mathrm{int}} H] \boldsymbol \rho \hat{P} + \hat{P}\boldsymbol\rho[\mathbf{Q},H \mathbf{M}_{\mathrm{int}}] (1- H \mathbf{M}_{\mathrm{int}})^{-1}\hat{P} + \notag \\ & + \hat{P} (1-\mathbf{M}_{\mathrm{int}} H)^{-1} [\mathbf{Q},\boldsymbol\mu] (1- H \mathbf{M}_{\mathrm{int}})^{-1}\hat{P}.
	\label{eq:Qrho}
\end{align*}
Note that $\mathbf{M}^2 = 0$ implies that $[\mathbf{Q},\mathbf{M}_{\mathrm{int}} ] = - \mathbf{M}_{\mathrm{int}}^2$ and that $H$ is a homotopy from $\mathbb{I}$ to $\hat{P}$, see (\ref{eq:hpt}). Therefore, the first and the second commutators yield
\begin{align}
	[\mathbf{Q}, \mathbf{M}_{\mathrm{int}} H] &= \mathbf{M}_{\mathrm{int}} \left(1 -  \mathbf{M}_{\mathrm{int}} H\right) - \mathbf{M}_{\mathrm{int}} \hat{P}, \\
	[\mathbf{Q}, H \mathbf{M}_{\mathrm{int}} ] &= - \left(1 -   H \mathbf{M}_{\mathrm{int}}\right) \mathbf{M}_{\mathrm{int}}  + \hat{P} \mathbf{M}_{\mathrm{int}}.
\end{align}
Using these results, we can simplify (\ref{eq:Qrho}) further and arrive at the identity
\begin{align*}
	[\mathbf{Q},{\boldsymbol\rho}] &= \hat{P} (1-\mathbf{M}_{\mathrm{int}} H)^{-1} [\mathbf{M},\boldsymbol\mu] (1- H \mathbf{M}_{\mathrm{int}})^{-1} \hat{P} = \frac{\partial}{\partial t} \mathcal{S}(\mathbf{M}).
\end{align*}
Using equation (\ref{eq:rho}) together with the gauge-invariance of the S-matrix (\ref{eq:Sgaugeinv}), we finally deduce a relation of the form
\begin{align}
	X \circ \frac{\partial}{\partial s} \mathcal{S}(\mathbf{M}) - \frac{\partial}{\partial t} \mathcal{S}(\mathbf{M}) &= -[\mathbf{Q},[\boldsymbol\eta,\xi\circ{\boldsymbol\rho}]],
\end{align}
from which the main equation (\ref{eq:main}) follows. This concludes the proof of equivalence of the superstring field theory proposed in \cite{SSFT1} with the ordinary perturbative string S-matrix for open superstrings in the NS-sector.

\section{Variations}\label{sec:app}
In the previous section we presented a proof of the equivalence of open superstring field theory to usual perturbative string theory in the NS-sector. However, the homological perturbation theoretical proof is applicable to some other, closely related physical systems: The action of  NS-NS sector of closed type II-superstring theory \cite{SSFT2} and the extension to the R-sectors at the level of the equations of motion \cite{SSFT3}. In both cases the construction is obtained by integrating the flow generated by an exact homological vector field on the formal manifold of homotopy algebraic structures. Now, in both cases the fundamental equation (\ref{eq:main}) still holds true,
\begin{align*}
	X \circ \frac{\partial}{\partial s} \mathcal{S}(\mathbf{M}) - \frac{\partial}{\partial t} \mathcal{S}(\mathbf{M}) &= [\boldsymbol\eta, [\mathbf{Q}, \cdots ] ].
	\label{eq:master}
\end{align*}
From the proof in section \ref{sec:proof} this follows quite trivially, because we only assumed that $\frac{\partial}{
\partial t} \mathbf{M} = [\mathbf{M}, \mathbf{S}]$ and that $\frac{\partial}{\partial s} \mathbf{M} = [\boldsymbol\eta, \mathbf{S}]$ for some $\mathbf{S}$.

\subsection{Closed type II-superstring}\label{sec:closed_ssft}
On the world-sheet of a closed type II superstring we have holomorphic and antiholomorphic degrees of freedom. On-shell NS-NS vertex operators in Siegel gauge take the form $V = V_m(z,\bar{z}) c(z) \bar{c}(\bar{z}) \delta(\gamma(z)) \delta(\bar{\gamma}(\bar{z}))$, where $V_m$ denotes a superconformal matter primary field of dimension $(\frac{1}{2},\frac{1}{2})$. The type II-worldsheet is a super Riemann surface with two odd directions and the world-sheet theory now comes with a holomorphic and an antiholomorphic picture number, both of which have to add up to $-2$ in order to obtain a non-vanishing correlator. In \cite{SSFT2} a holomorphic and an antiholomorphic picture deficit together with formal variables $s$ and $\bar{s}$ were introduced. A generic coderivation $\mathbf{L}$ can then be identified with
\begin{align*}
	\mathbf{L} &= \sum_{k,l \geq 0} s^k \bar{s}^l \mathbf{L}^{[k,l]},
\end{align*}
where $\mathbf{L}^{[k,l]}$ has holomorphic picture deficit $k$ and antiholomorphic picture deficit $l$.

Closed string products are graded-symmetric, hence, the underlying homotopy algebraic structure is an $L_\infty$ algebra instead of an $A_\infty$-algebra. However, it is possible to take the universal envelope of an $L_\infty$-algebra \cite{Lada:1994mn} and obtain an $A_\infty$-algebra to which the usual construction can be applied. Alternatively, one can think of the construction in the dual geometric picture and skip the universal enveloping algebra completely. Eventually, two vector fields $\delta$ and $\bar{\delta}$ were introduced,
\begin{align}
	\delta \mathbf{L} &= [\mathbf{L}, \boldsymbol{\lambda} ] & \bar{\delta} \mathbf{L} &= [\mathbf{L}, \bar{\boldsymbol{\lambda}} ] \\
	[\boldsymbol\eta, \boldsymbol{\lambda}] &= \frac{\partial}{\partial s} \mathbf{L} & [\bar{\boldsymbol\eta}, \bar{\boldsymbol\lambda} ] &= \frac{\partial}{\partial \bar{s}} \mathbf{L}. \label{eq:lambdadef}
\end{align}
The equations (\ref{eq:lambdadef}) were then solved using the special contracting homotopy for $\eta$ or $\bar{\eta}$ that was built using the zero-modes of the $\xi$- or $\bar{\xi}$-fields. This was required to preserve the constraint $b_0^- = L_0^- = 0$ on the closed-string state space. However, in the following we do not require this choice for $\boldsymbol\lambda$ and $\bar{\boldsymbol\lambda}$.

The closed string products are then obtained by integrating the flow of (some, possibly $t$-dependent) linear combination of $\delta$ and $\bar{\delta}$ starting at a point where $\mathbf{L} = \mathbf{L}_{\mathrm{bos}}$ coincides with the closed string vertices of closed bosonic string field theory \cite{Zwiebach:1992ie} and the so-obtained closed string products are all related by a field redefinition in the small Hilbert space\footnote{This can be seen by showing that the commutator $[\delta,\bar{\delta}]$ is a field redefinition. In fact, $[\delta,\bar{\delta}] \mathbf{L} = [\mathbf{L},\boldsymbol\kappa]$ with $\boldsymbol\kappa = [\boldsymbol\lambda,\bar{\boldsymbol\lambda}] + \bar{\xi}\circ\frac{\partial}{\partial \bar{s}} [\mathbf{L},\boldsymbol\lambda] - \xi\circ\frac{\partial}{\partial s} [\mathbf{L},\bar{\boldsymbol\lambda}] - [\mathbf{L},\xi\circ\bar{\xi}\circ\frac{\partial^2}{\partial \bar{s} \partial s}\mathbf{L}]$ and $[\boldsymbol\eta,\boldsymbol\kappa] = [\bar{\boldsymbol\eta},\boldsymbol\kappa] = 0$. Thus, on the space of $L_\infty$-structures modulo field redefinitions $[\delta,\bar{\delta}] = 0$ and the endpoint of the flow of $a \delta + b \bar{\delta}$ depends only on the duration in $\delta$ or $\bar{\delta}$-direction.}. It therefore suffices to consider the special case for which we integrate
\begin{align*}
	\frac{\partial}{\partial t} = \Delta &= \delta + \bar{\delta}. 
\end{align*}
Let us remark that our argument also works for arbitrary $t$-dependent linear combinations for $\Delta$, but for simplicity we restrict to this special choice.
The main equation (\ref{eq:master}) now tells us that
\begin{align}
	X \circ \mathcal{S}(\mathbf{L}) + \bar{X} \circ \mathcal{S}(\mathbf{L}) = \frac{\partial}{\partial t} \mathcal{S}(\mathbf{L}) + [\boldsymbol\eta, [\bar{\boldsymbol\eta}, [\mathbf{Q}, \cdots]]].
	\label{eq:closedSmatrix}
\end{align}
The rest of the argument is very similar to the one given in section \ref{sec:proof}. We will only work out the details for the four-point S-matrix elements here. The closed string S-matrix elements are calculated from $\mathcal{S}(\mathbf{L})$ using the symplectic form $\omega_S = \omega( \mathbb{I}\otimes \xi_0\bar{\xi}_0 c_0^-)$, where $\omega$ denotes the BPZ-inner product for the world-sheet theory formulated in the large Hilbert space. The S-matrix is then the restriction of the functional $S$
\begin{align*}
	S &= \omega_S( \mathbb{I} \otimes \mathcal{S}(\mathbf{L}) )
\end{align*}
to the relative cohomology $H^\bullet(Q|\eta,\bar{\eta})$. Equation (\ref{eq:closedSmatrix}) decomposes into a system of differential equations in the deformation parameter $t$ by reading off the coefficients of terms homogeneous in $s$ and $\bar{s}$:
\begin{align*}
	\frac{\partial}{\partial t} S^{[0,0]}_4(t) &= S^{[1,0]}_4(t) X + S^{[0,1]}_4(t) \bar{X} \\
	\frac{\partial}{\partial t} S^{[1,0]}_4(t) &= 2 S^{[2,0]}_4(t) X + S^{[1,1]}_4(t) \bar{X} \\
	\frac{\partial}{\partial t} S^{[0,1]}_4(t) &= S^{[1,1]}_4(t) X + 2 S^{[0,2]}_4(t) \bar{X} \\
	\frac{\partial}{\partial t} S^{[1,1]}_4(t) &= 2 S^{[2,1]}_4(t) X + 2 S^{[1,2]}_4(t) \bar{X} \\
	\frac{\partial}{\partial t} S^{[2,0]}_4(t) &= S^{[2,1]}_4(t) \bar{X} \\
	\frac{\partial}{\partial t} S^{[0,2]}_4(t) &= S^{[1,2]}_4(t) X \\
	\frac{\partial}{\partial t} S^{[2,1]}_4(t) &= 2 S^{[2,2]}_4(t) \bar{X} \\
	\frac{\partial}{\partial t} S^{[1,2]}_4(t) &= 2 S^{[2,2]}_4(t) X \\
	\frac{\partial}{\partial t} S^{[2,2]}_4(t) &=0.
\end{align*}
In the last equation we used the fact that the highest picture deficit for $3$-products in this construction is $[2,2]$ so that there are no source terms of the last differential equation. Indeed, the functional $S^{[2,2]}_4$ is identical to the $S$-matrix calculated from  bosonic CSFT described by the initial vertices. It is clear that this system of equations can be integrated directly and we can express the $S$-matrix $S^{[0,0]}_4$ in terms of the bosonic CSFT-S-matrix $S_{\mathrm{bos},4} = S^{[2,2]}_4(0)$ and picture changing operators $X$ and $\bar{X}$ located at the punctures,
\begin{align*}
	S^{[0,0]}_4 &= S_{\mathrm{bos},4} X^2 \bar{X}^2.
\end{align*}
Moreover, if the external states are on-shell, we can move the PCOs arbitarily and may adjust them such that all external states are in the $(0,0)$ picture except for two that are in the $(-1,-1)$ picture.

\subsection{Equations of motion for the Ramond fields}\label{sec:ramond_ssft}
Formulating the dynamics of the Ramond string fields in the small Hilbert space using an action principle is still an open problem. Finding covariant equations of motion is a somewhat simpler problem and was solved recently using homotopy algebraic methods \cite{SSFT3}. In this subsection we briefly describe this construction but using a more condensed notation that makes the connection to the methods used here more clearly without obscuring the overall picture by details. Furthermore, we only discuss the validity of the resulting equations of motion for the open superstring obtained from the stubified bosonic open string products. The extension to the closed type II superstring and the heterotic string contains no new conceptual ideas and we leave the details to the enthusiastic reader.

The string field $\phi = \phi_{\mathrm{NS}} + \phi_{\mathrm{R}}$ now takes values in the CFT state space $\mathcal{H}_{\mathrm{NS}} \oplus \mathcal{H}_{\mathrm{R}}$, where the NS field is at picture $-1$ and the R field is at picture $-\frac{1}{2}$. The final result of the construction of \cite{SSFT3} is a homological vector field on the non-commutative manifold whose function ring is given by $T\mathcal{H}^*$. The zeros of this vector field coincide with the solutions to the equations of motions. Let us recall that if this vector field came from an action $S$, then it could be written as a normal vector to the surface $S = \mathrm{const.}$ using a symplectic form to identify tangent with cotangent vectors. However, no consistent truncation of the open string state space is known so that the BPZ inner product reduces to a symplectic form in the R-directions. If such a product would exist, then one could try to find an integrating factor such that the vector field actually becomes a gradient vector field of some function. However, even if no action for a set of equations of motion is known, there are still some questions that can be answered. For example, one can ask about the structure of the space of solutions modulo gauge transformations. The cotangent complex of such an algebraic variety is characterized by the classical S-matrix, when the equations of motion arise from an action. Thus, we can still calculate the S-matrix for the R-equations of motion. Equality to the S-matrix of perturbative string theory is a necessary criterion for calling the equations from \cite{SSFT3} equations of motions for the open superstring as the universal deformations should coincide. In \cite{Sen:2015hha} it was shown that one may calculate the classical S-matrix directly from a pseudo-action.
The obstruction to the existence of smooth directions in the moduli space of solutions is measured by the minimal model of the homological vector field \cite{Cattaneo:2012qu}. The minimal model is an $A_\infty$ structure on $H^\bullet(Q|\eta)$ that is obtained by restricting $\mathcal{S}(\mathbf{M})$ to the cohomology. We may contract the minimal model structure with a non-degenerate symplectic form on $H^\bullet(Q|\eta)$ to obtain a linear functional $S$, which is the classical S-matrix of the equations of motion. In order to define said symplectic structure, we need to introduce an inverse picture changing operators $Y$ that is required to be BPZ-even and a homotopy inverse of $X$. We now introduce an operator $\mathcal{O}$ by
\begin{align*}
	\mathcal{O} \phi &= \phi_{\mathrm{NS}} + Y \phi_{\mathrm{R}}.
\end{align*}
The sought for symplectic form $\tilde{\omega}$ is now in terms of the large Hilbert space BPZ-inner product $\omega$,
\begin{align*}
	\tilde{\omega} &= \omega(\mathbb{I} \otimes \xi \mathcal{O}).
\end{align*}
It is readily checked that $\tilde{\omega}$ is $Q$-closed and, hence, descends to a non-degenerate pairing on $H^\bullet(Q|\eta)$. The S-matrix for a homological vector field $\mathbf{M}$ is then
\begin{align}
	S &= \tilde{\omega}(\mathbb{I} \otimes \pi_1\mathcal{S}(\mathbf{M})).
	\label{eq:SmatrixR}
\end{align}
The main difference to the construction for the pure NS-subsector is that we now have two component fields $\phi_{\mathrm{NS}}$ and $\phi_{\mathrm{R}}$ which carry different picture number. Thus, the required number of PCO insertions will depend on the sector of the inputs to a vertex. In general, if we substitute a pair of NS string fields with a pair of R string fields, the number of required PCOs reduces by one. In order to keep track of these requirements, in \cite{SSFT3} a new auxiliary quantum number, \emph{Ramond number}, was introduced and the corresponding formal variables was called $u$, i.e.~all coderivations are formal power series in $u$ with coefficients of $u^r$ having Ramond number $r$. A product $M_n$ has Ramond number $r$ if and only if it vanishes unless it has $2r$ or $2r+1$ number of Ramond inputs. Moreover, the notation $M_n|_{2r}$ was introduced to denote the restriction to Ramond number $r$. Notice that Ramond number is additive under taking commutators of coderivations so that $u$ indeed counts Ramond number.

Denote now by $N_k$ the bosonic open string products and define a codifferential $\mathbf{M}(0)$
\begin{align*}
	\mathbf{M}(0) &= \mathbf{Q} + u \mathbf{N}_2 |_{2} + s \mathbf{N}_2|_{0} + s^2 ( \mathbf{N}_3|_{0} + u \mathbf{N}_4|_{2} ) + \cdots \\
		&= \sum_{n=0,r=0}^\infty s^n u^r \mathbf{N}_{n+r+1}|_{2r}.
\end{align*}
We use $\mathbf{M}(0)$ as a starting point for the usual deformation equations (\ref{eq:def1},\ref{eq:def2}) and integrate the flow. The equations of motion are then the Maurer-Cartan equations for the codifferential $\mathbf{M}$ at the end of the flow, when restricted to picture deficit $0$. By construction the products in $\mathbf{M}^{[0]}$ produce string fields in picture $-1$ or $-\frac{1}{2}$. We are now ready to evaluate the S-matrix (\ref{eq:SmatrixR}) in the same way as in section \ref{sec:proof}. The functional $S$ is then equal to the bosonic S-matrix with vertex operators inserted in the correct picture if the output of $\mathcal{S}(\mathbf{M})$ is an NS-state. If it is an R-state, we can use one of the PCOs to remove the $Y$ operator at the output and we still obtain the perturbative string S-matrix. Let us see how this works for the four-point amplitude of two $R$-states $R_1$ and $R_2$ with two $NS$-states $NS_1$ and $NS_2$. The relevant component of $S_4^{[0]}$ has Ramond number $0$ and is given in terms of the bosonic S-matrix $\mathcal{S}(\mathbf{M}_{\mathrm{bos}})$ as
\begin{align*}
	S_4^{[0]}(R_1,R_2,NS_1,NS_2) &= \omega( R_1 \otimes \xi Y (X\circ X \circ \mathcal{S}(\mathbf{M}_{\mathrm{bos}}) (R_2,NS_1,NS_2)) = \\
	&= \omega( X Y R_1 \otimes \xi \mathcal{S}(\mathbf{M}_{\mathrm{bos}})(X R_2, NS_1, NS_2)) = S_{\mathrm{bos}}(R_1,X R_2, NS_1, NS_2).
\end{align*}
This concludes our discussion of the validity of the construction in \cite{SSFT3} as valid Ramond equations of motion.

\subsection{Relation to Berkovits' WZW-like theory}\label{sec:berkovits}
Our result has further implications. In \cite{Erler:2015rra} it was shown that the CS-like formulation of open super string field theory from \cite{SSFT1} is related to a gauge-fixed version of Berkovits' WZW-type super string field theory through a field redefinition. Now, since the S-matrix is invariant under field redefinitions up to a similarity transformation, our result states that the S-matrix of Berkovits' WZW-type formulation agrees with the usual perturbative super string S-matrix. Previously \cite{Berkovits:1999bs, Iimori:2013kha, Kunitomo:2015hda} some checks in this direction were performed, but remained restricted to the four-point and five-point S-matrix elements. Equivalence of CS-like heterotic string field theory and its WZW-like formulation has been studied recently in \cite{Goto:2015hpa}.

\section{Conclusions}
We showed the equivalence of perturbative string theory with the super string field theories based on the small Hilbert space. This equivalence requires that the solution space of the linearized equations of motion coincides with the physical string spectrum and that the S-matrix around a given vacuum agrees with that of perturbative string theory. The first requirement was true by construction and we only had to show the second. In doing so, the special form of the cohomological vector field encoding the equations of motion was crucial: It allowed us to relate the S-matrix of the underlying bosonic string field theory to the real S-matrix by a sequence of descent equations (\ref{eq:descent}) without employing complicated combinatorial arguments involving Feynman diagrams or worldsheet diagrams.

Despite the progress at the algebraic level, there are still some open questions to address. For example, it would be interesting to see if and how the algebraic construction and properties arise from the world-sheet point of view. Since the formulation is entirely in terms of the small Hilbert space expressing the interaction vertices in terms of integrals over the moduli space of super Riemann surfaces should be easier than in the large Hilbert space formulations. However, even though formulated in terms of small Hilbert space fields, we still use the bosonized $\beta$-$\gamma$-ghosts. A first step towards a geometric formulation would be to reformulate the construction in terms of operators manifestly built from modes of the $\beta$ and $\gamma$-ghosts and to find a geometric interpretation of the descent equations.
Quantization of the theory necessitates an action principle. However, since only equations of motion are known for the Ramond string fields, the first step must be to reformulate them in terms of an action principle. In turn we would need to find a suitable symplectic form of picture number $-1$ on the space of (off-shell) Ramond fields. Most likely such a construction would require a constraint on the Hilbert space, similar to the construction of closed string field theory. Some proposals \cite{Jurco:2013qra,Kohriki:2012pp} in this direction exist, but it is not clear if these constraints do not alter the spectrum of physical states.

\paragraph{Acknowledgements} The author would like to thank Ted Erler for critically reading the manuscript and Ivo Sachs for helpful comments. This project was supported in parts by the DFG Transregional Collaborative Research Centre TRR 33, the DFG cluster of excellence Origin and Structure of the Universe.

\appendix
\section{Strong deformation retracts and cohomomorphisms}\label{app:coh}
In this appendix we want to elaborate on some claims made in section \ref{sec:hpt} according which the perturbed projections $\mathfrak{p}'$ and inclusions $\mathfrak{i}'$ are cohomomorphisms and the perturbed differential $D'$ is actually a coderivation. We claimed that the property $p Q^\dagger = Q^\dagger i = (Q^\dagger)^2 =  0$, which implies that we are working with a strong deformation retract, is sufficient.

The key equations of this proof are the following two identities,
\begin{align*}
	\left( ( h \otimes' \mathfrak{i}\mathfrak{p} + \mathbb{I} \otimes' h)( \delta \otimes' \mathbb{I} + \mathbb{I} \otimes' \delta ) \right)^k (\mathfrak{i} \otimes' \mathfrak{i}) &= \sum_{r+s = k, r,s \geq 0} ((h\delta)^r \otimes' (h\delta)^s)(\mathfrak{i} \otimes' \mathfrak{i}), \\
	(\mathfrak{p}\otimes' \mathfrak{p}) \left( ( h \otimes' \mathfrak{i}\mathfrak{p} + \mathbb{I} \otimes' h)( \delta \otimes' \mathbb{I} + \mathbb{I} \otimes' \delta ) \right)^k&= \sum_{r+s = k, r,s \geq 0}  (\mathfrak{p} \otimes' \mathfrak{p})  ((h\delta)^r \otimes' (h\delta)^s).
\end{align*}
This equation can be proven by mathematical induction. The proof in both cases is very similar, so that we only sketch it for the first identity. Indeed, the case $k = 0$ is obvious. So suppose that the above equation is true for some $k \geq 0$. Then it follows that
\begin{align*}
	\left( ( h \otimes' \mathfrak{i}\mathfrak{p} + \mathbb{I} \otimes' h)( \delta \otimes' \mathbb{I} + \mathbb{I} \otimes' \delta ) \right)^{k+1} (\mathfrak{i} \otimes' \mathfrak{i}) &= \sum_{r+s = k, r,s \geq 0} ( h\delta \otimes' \mathfrak{ip} + h \otimes' \mathfrak{ip}\delta - \delta \otimes' h + \mathbb{I} \otimes' h\delta ) ((h\delta)^r \otimes' (h\delta)^s)(\mathfrak{i} \otimes' \mathfrak{i}) \notag \\
	&= \sum_{r+s = k, r,s \geq 0}  ( \delta_{s,0} (h\delta)^{r+1} \otimes' (h\delta)^s + (h\delta)^r \otimes' (h\delta)^{s+1}))(\mathfrak{i} \otimes' \mathfrak{i}) \notag \\
	&= \sum_{r+s = k+1, r,s \geq 0}  ( (h\delta)^r \otimes' (h\delta)^{s})(\mathfrak{i} \otimes' \mathfrak{i}).
\end{align*}
This concludes the proof. One immediate corollary is that $\mathfrak{i}'$ is a cohomomorphism,
\begin{align*}
	\Delta \mathfrak{i}' &= \Delta (1 - h \delta )^{-1} \mathfrak{i} = \sum_{k=0}^\infty \left( ( h \otimes' \mathfrak{i}\mathfrak{p} + \mathbb{I} \otimes' h)( \delta \otimes' \mathbb{I} + \mathbb{I} \otimes' \delta ) \right)^k (\mathfrak{i} \otimes' \mathfrak{i}) \Delta = (\mathfrak{i}' \otimes' \mathfrak{i}' ) \Delta.
\end{align*}
Similarly, one can prove that $\mathfrak{p}'$ is a cohomomorphism. One further simple consequence is that $D'$ is a coderivation,
\begin{align*}
	\Delta D' &= (D \otimes' \mathbb{I} + \mathbb{I} \otimes' D) \Delta + (\mathfrak{p}\otimes'\mathfrak{p}) ( \delta \otimes' \mathbb{I} + \mathbb{I} \otimes' \delta) (\mathfrak{i}'\otimes\mathfrak{i'}) \Delta = (D' \otimes' \mathbb{I} + \mathbb{I} \otimes' D') \Delta. 
\end{align*}

\bibliography{smatrix}
\bibliographystyle{utphys}
\end{document}